# Deep Learning Superresolution for 7T Knee MR Imaging: Impact on Image Quality and Diagnostic Performance

*Pinzhen Chen\*, PhD*● Libo *Xu\*, MS*● Boyang *Pan, MS*● *Jing Li, MS*● Yuting *Wang, MS*● Ran *Xiong, MD*● Xiaoli *Gou, MS*● Long *Qing, PhD*● *Wenjing Hou, MS*● Nan-jie Gong*, PhD*● *Wei Chen, MD*

From 7T Magnetic Resonance Translational Medicine Research Center, the Departments of Radiology (P.Z.C., J.L.,Y.T.W., W.J.H, W.C.) and Sports Medicine (R. X., X.L.G), Southwest Hospital, Army Medical University (Third Military Medical University), Chongqing, China.

From the MR Research Collaboration Team (L.Q, PhD), Siemens Healthineers Ltd., Guangzhou, China.

RadioDynamic Healthcare (L.B.X, B.Y.P), Shanghai, China.

Laboratory for Intelligent Medical Imaging (N.J.G), Tsinghua Cross-Strait Research Institute, Tsinghua University, Beijing, China

Co-first authors: *These authors contributed equally to this work.

Address correspondence to *Wei Chen, MD* (e-mail: landcw@tmmu.edu.cn )
7T Magnetic Resonance Translational Medicine Research Center, Department of Radiology, Southwest Hospital, Army Medical University (Third Military Medical University), 30 Gaotanyan Street Shapingba District, Chongqing, 400038, China.

Supported by Chong Qing Natural Science Foundation(CSTB2023NSCQ-MSX0919).

Manuscript Type: Original Research
Word Count for Text: 3576
Data sharing statement: Data generated or analyzed during the study are available from the corresponding author by request.

**Abstract**

**Background:** Deep learning based superresolution has the potential to enhance the image quality of musculoskeletal MR imaging; however, its diagnostic value in knee imaging at 7 T remains uncertain.

**Objectives:** To evaluate image quality and diagnostic performance of deep learning superresolution (SR) 7T knee MR images compared with low-resolution (LR) and high-resolution (HR) acquisitions.

**Materials and Methods:**

This prospective study enrolled 42 participants (54 knees; mean age, 45 years ± 17; 18 men) who underwent 7T knee MRI with LR (0.8 × 0.8 × 2 mm³) and HR (0.4 × 0.4 × 2 mm³) proton density turbo spin-echo fat-suppressed sequences. A Hybrid Attention Transformer model generated SR images from LR input. Three radiologists independently assessed image quality and anatomic conspicuity using five-point Likert scales. Detection of cartilage lesions, meniscal tears, ligament tears, and bone marrow abnormalities was recorded. Arthroscopy served as the reference standard in 10 participants. Statistical analyses included Gwet AC2 coefficient, Friedman, McNemar, Cohen κ, and diagnostic performance tests.

**Results:** SR images scored higher for overall image quality than LR images (median score 5 vs 4; $P < .001$), and exhibited lower noise than HR images (median score 5 vs 4; $P < .001$). Visibility of cartilage, menisci, and ligaments was superior in SR and HR images (median score: 5) compared to LR images (median score: 4) ($P < .001$). The tibial nerve was best seen in HR images (score: 4), followed by SR (3.5) and then LR (3). The detection frequencies and diagnostic performance metrics (sensitivity, specificity, area under the curve) for intra-articular pathology are similar among three types images ($P \geq 0.095$).

**Conclusions:** Deep learning superresolution improved image quality of 7T knee MR images but did not enhance diagnostic performance for detection of intra-articular pathology compared with standard LR imaging.

**Abbreviations**

Area under the curve: AUC

Convolutional Block Attention Module: CBAM

Hybrid Attention Transformer: HAT

High-resolution: HR

Low-resolution: LR

Superresolution: SR

Overlapping Cross-Attention Block: OCAB

Proton Density Turbo Spin Echo fat suppression: PD-TSE FS

Residual Hybrid Attention Group: RHAG

**Summary Statement**

Deep learning superresolution improved 7T knee MR image quality but did not enhance diagnostic performance for intra-articular pathology compared to standard low-resolution imaging.

**Key Results**

In this prospective study of 42 participants (54 knees) undergoing 7T knee MRI, superresolution (SR) images demonstrated significantly superior overall image quality and visibility of knee structures than low-resolution LR images (P < .001).
Diagnostic performance for intra-articular pathology was similar between SR, LR, and HR images, with no significant differences in sensitivity (e.g., LR:62% vs. SR and HR: 69% for cartilage defects) or specificity (LR and SR: 98% vs. HR: 98%) (P ≥ .095).

**Introduction**

Knee disorders are among the most frequent musculoskeletal conditions worldwide and a leading cause of pain, disability, and healthcare utilization (1). Magnetic resonance imaging (MRI) is the modality of choice for evaluating internal derangements of the knee due to its superior soft-tissue contrast and multiplanar capability. However, standard knee MR protocols at 1.5T or 3T are often limited by relatively low in-plane resolution, which can restrict visualization of subtle morphologic changes in cartilage, menisci, and ligaments.

Ultra-high-field 7T MR imaging offers an increased signal-to-noise ratio and permits higher spatial resolution compared to lower-field-strength systems. These improvements enable a more detailed depiction of joint anatomy(2); however, the widespread clinical adoption of 7T MR imaging has been limited by prolonged scan times, increased susceptibility to motion and artifacts, and higher costs. Techniques such as parallel imaging (3, 4), simultaneous multislice acquisition (5, 6), and compressed sensing (7) have reduced acquisition times and improved image quality in musculoskeletal 7T MRI for healthy adults. Nervertheless, developing further strategies to improve resolution without extending scan duration remains clinically important for diagnosing knee joint disorders.

Deep learning–based image reconstruction has emerged as a promising solution. By learning mappings between low- and high-resolution data, deep learning super-resolution models can generate high-resolution (HR) images from faster low-resolution (LR) acquisitions without requiring hardware modifications. This approach may improve spatial resolution and visualization of delicate anatomic structures while maintaining clinically feasible scan times (8). Convolutional neural networks (CNNs) and generative adversarial networks (GANs) are the most commonly applied architectures. CNN-based models produce smooth images but risk loss of fine textural details (9, 10), while GANs better preserve natural appearance but are prone to training instability and hallucinated features (11, 12).

The Hybrid Attention Transformer (HAT) is a recently developed deep learning model designed to overcome these limitations. By integrating channel attention, window-based self-attention schemes, and an overlapping cross-attention module, HAT algorithm can foster the interaction among adjacent window features while maintaining training stability [21]. HAT has demonstrated superior performance in image restoration tasks such as brain MR imaging superresolution (13). However, its application to musculoskeletal imaging, particularly 7T knee MR imaging, remains unexplored.

Previous work has shown that deep learning reconstructions can enhance quantitative image quality in 7T knee MR imaging of healthy volunteers (14, 15).Yet the clinical impact of these methods for detecting intra-articular pathologies is unknown. Improvements in image quality do not necessarily translate to better diagnostic accuracy, and the value of deep learning super resolution for knee imaging in symptomatic patients has not been established.

The purpose of this study was to evaluate the performance of a deep learning superresolution model (HAT algorithm) in 7T knee MR imaging. Specifically, we sought to (a) compare image quality of superresolution images with standard low- and high-resolution acquisitions, and (b) assess whether superresolution improves diagnostic performance for detecting intra-articular abnormalities, using arthroscopy as a reference standard in a subset of participants.

**Method**

**Study Sample**

The acquisition of MRI data was approved by the institutional review board (KY2022147). All participants provided written informed consent. Between December 2022 and April 2024, 42 participants with 54 knees (32 left, 22 right knees) referred for diagnostic knee MRI were enrolled. Ten participants of the study sample underwent arthroscopic surgery. The selection criteria are as follows: a) age $\geqslant 18$ years old, b) having no metal implants or contraindications, and c) being able to fit in the knee coil or MRI scanner. Participants were evaluated for cartilage abnormalities, meniscal and ligament injuries, as well as bone marrow abnormalities.

**MRI Protocol**

Knee data were prospectively acquired using a 7T MR system (MAGNETOM Terra,

Siemens Healthineers) for the axial, coronal, and sagittal Proton Density Turbo Spin Echo fat suppression (PD-TSE FS) sequence scanning. The 7T MR knee paired data included 42 subjects who underwent both LR (0.8 mm in-plane resolution) and HR (0.4 mm in-plane resolution) examinations prospectively. **Table 1** provides detailed scan parameters.

**Deep Learning Method**

We propose an enhanced framework based on the HAT (Hybrid Attention Transformer) algorithm(16, 17) for 7T MR joint image super-resolution. Several key modifications to the original HAT algorithm were implemented to optimize its performance in medical imaging: 1) The conventional convolution layers is replaced with a multi-branch Inception-style structure to promote multi-scale feature fusion and reduce parameter complexity. 2) A Convolutional Block Attention Module (CBAM) is integrated to incorporate both channel and spatial attention, enabling the model to better capture critical spatial information. For the 7T MR knee images SR task, we used an mean squared error (MSE)-based HAT and then adopted matched 7T knee datasets with in-plane resolutions of 0.8mm and 0.4mm for training (as shown in **Figure 1**).

The dataset consisted of 54 knee MRI pairs. We allocated 24 pairs for model training and reserved the remaining 30 for testing. Additionally, the 10 participants who underwent arthroscopic surgery were excluded from training. This ensured a truly independent test set to avoid data leakage, thereby ensuring the model learned a reliable LR-to-HR mapping.

**Network structure of the enhanced framework**

To adapt HAT for 7T knee MRI super-resolution, our network retains the classic Residual in Residual architecture of the original HAT but replaces all conventional convolution layers with multi-branch Inception-style structures (1×1, 3×3, 5×5 convolutions in parallel) [17] and CBAM for diagnostic-region prioritization [18]. This design optimizes for 7T knee MRI characteristics and ensures multi-scale capture of anatomical features (from micro cartilage defects to macro ligament morphology). The architecture consists of three tailored modules: shallow feature extraction, deep feature extraction, and image reconstruction.

**Shallow Feature Extraction (Inception-based)**

Unlike the original HAT that uses a single 3×3 convolution for shallow feature extraction, we replace this layer with an Inception-style convolution layer (Fig. 1(a) "Input image → Residual Hybrid Attention Groups (RHAG)" stage) to simultaneously capture heterogeneous knee MRI features: 1×1 convolution to reduce channel dimensionality to avoid computational redundancy, adapting to 384×384 matrix's high pixel count; 3×3 convolution to capture local fine details critical for detecting grade 2A/2B cartilage defects; 5×5 convolution to captures global structural features to preserve anatomical integrity.

Specifically, given an input LR 7T knee MRI image $I_{LR} \in \mathbb{R}^{H \times W \times C_{in}}$, we apply an Inception-based shallow feature extractor $H_{Inception-SF}(\cdot)$ to extract high-dimensional shallow features $F_0 \in \mathbb{R}^{H \times W \times C}$ as follows:

$$F_0 = H_{Inception-SF}(I_{LR}) \#(1)$$

where $C_{in}$ (input channels, 1 for PD-TSE FS) and $C$ (intermediate channels, 144) denote the channel count of input and feature maps, respectively. This Inception-based design not only maps input from low-dimensional to high-dimensional space but also generates pixel-level embeddings that retain both pathological signals and anatomical context [20], laying the foundation for subsequent deep feature refinement.

**Deep Feature Extraction (Inception + CBAM Enhancement)**

The deep feature extraction module $H_{DF}(\cdot)$ (Fig. 1(a) "RHAG" blocks) further refines $F_0$ into diagnostic-oriented deep features $F_{DF} \in \mathbb{R}^{H \times W \times C}$, computed as:

$$F_{DF} = H_{DF}(F_0) \#(2)$$

where $H_{DF}(\cdot)$ comprises 6 RHAG and one Inception-style convolution layer. Each RHAG (Fig. 1(b)) consists of 6 Hybrid Attention Blocks (HABs), one Overlapping Cross-Attention Block (OCAB, Fig. 1(d)), and one Inception-style convolution layer—all connected via a residual connection to avoid gradient vanishing. The RHAGs sequentially refine intermediate features through multi-scale attention fusion:

$$F_i = H_{RHAG_i}(F_{i-1}), i = 1,2, ..., N,$$
$$F_{DF} = H_{Inception-Conv}(F_N) \#(3)$$

where $H_{RHAG_i}(\cdot)$ denotes the $i-th$ RHAG, and $H_{Inception-Conv}(\cdot)$ is the Inception-style convolution layer.

Each HAB (Fig. 1(c)) retains the original window-based multi-head self-attention ((S)W-MSA) but replaces the channel-only Attention Block (CAB) in HAT with CBAM (Fig. 1(e))—a dual-attention module that combines channel and spatial attention to prioritize diagnostic-relevant features: channel attention adaptively weights feature channels based on signal intensity, highlighting high-signal regions; spatial attention generates a spatial weight map to focus on anatomically critical regions while suppressing non-target tissues. For an input feature X of HAB, the computation (integrating CBAM) is formulated as:

$$X_N = LN(X),$$
$$X_M = (S)W - MSA(X_N) + \alpha CBAM(X_N) + X, \#(5)$$
$$Y = MLP(LN(X_M)) + X_M,$$

where $X_N$ and $X_M$ are intermediate features, and $Y$ is the HAB output. $LN(\cdot)$ represents the layer normalization, and MLP is a standard multi-layer perceptron. (S)W-MSA refers to either standard or shifted window-based multi-head self-attention. A small constant $\alpha = 0.01$ is set to balance (S)W-MSA and CBAM contributions, avoiding attention dominance.

Retaining the original HAT design, OCAB adopts overlapping window strategies for cross-window feature interaction: it partitions query features into non-overlapping windows while using overlapping windows for key/value projections, which effectively strengthens feature connections between adjacent windows and avoids isolated local feature extraction. This design is critical for preserving the continuity of knee anatomical structures in 7T MRI images, and complements the multi-scale fusion capability of Inception-style convolutions and the diagnostic-region focus of CBAM.

**Image Reconstruction (Inception-Aided Fidelity Preservation)**

Following (17), an additional Inception-style convolution layer is inserted at the end of this stage to better aggregate deep features. Next, to generate HR 7T knee MRI images, we first fuse shallow features $F_0$ and deep features $F_{DF}$ via a global residual connection (to retain initial anatomical context), then input the fused features into an Inception-enhanced reconstruction module $H_{Inception-Rec}(\cdot)$ (Fig. 1(a) "Output ×4 image" stage):

$$I_{HR} = H_{Inception-Rec}(F_0 + F_{DF}) \#(4)$$

where $H_{Inception-Rec}$ is composed of an Inception-style convolution layer, a pixel-shuffle up-sampling with upscale factor=4, and another Inception-style convolution layer. This Inception-aided reconstruction design avoids over-smoothing of diagnostic details and ensures the final SR images retain the high-fidelity required for clinical interpretation.

**Image Quality and the visibility of knee anatomical structure.**

Three musculoskeletal radiologists independently assessed the images on the following metrics by rating them on a five-point Likert scale: overall image quality, motion artifacts, Gibbs artifacts, and image noise. Visibility of the articular cartilage, medial and lateral meniscus, anterior and posterior cruciate ligament, medial and lateral collateral ligament, and tibial nerve was also assessed using a five-point equidistant Likert scale. Definitions of Likert ratings are as follows: 1 (very bad/severe); 2 (bad/moderate); 3 (adequate/mild); 4 (good/minimal); and 5 (very good/absent). The results from the three readers were averaged for further statistical analysis.

**Detection of Structural Abnormalities**

Articular cartilage defects were assessed by compartment, including the medial and lateral femur, medial and lateral tibia, patella, and trochlea, and graded using a five-category grading scheme based on the modified Noyes classification. Definitions of Noyes classification are as follows: grade 0 (normal cartilage), grade 1 (increased T2 signal intensity of morphologically normal cartilage not oriented at 55° to the external magnetic field), grade 2A (superficial partial-thickness cartilage defect <50% of the total articular surface thickness), grade 2B (deep partial-thickness cartilage defect >50% of the total articular surface thickness), and grade 3 (full-thickness cartilage defect) (18). In cases where multiple defects were identified within a single compartment, only the dominate lesion was taken into consideration.

Tears of the menisci was rated binary ("absent" vs. "present"). Proton density hyperintense signal breaching a meniscus articular surface was defined as a tear. The presence of abnormalities was rated as binary ("absent" vs. "present") for the anterior and posterior cruciate ligaments, as well as the medial and lateral collateral ligaments. Ligaments were considered torn when MRI showed a greater than 50% cross-sectional fiber discontinuity.

The detection frequency of structural abnormalities was quantified by presenting both the numerator and denominator, along with corresponding percentages, for each reader. Additionally, a consensus frequency was established based on majority agreement among the three independent readers.

**Evaluation for Hallucinations and Spurious Omissions**

Three musculoskeletal radiologists reviewed results and MRI scans independently for spurious creations and omissions of abnormalities on DL super-resolution MRI scans.

**Statistical Analysis**

Likert scores are provided as median values with IQRs. The inter-rater agreement among the three readers for qualitative measurements was assessed using Gwet AC2 coefficient with linear weights for binary data and quadratic weights for articular cartilage defects data and quality scores (19, 20).

The differences in qualitative measurements resulting from the five-point Likert scale across three types of images were evaluated using the Friedman test, followed by the pairwise post hoc Wilcoxon test with Bonferroni-Holm correction to correct for multiple comparisons. Cohen's kappa (κ) and McNemar's test were used to quantify intermethod agreement and to compare diagnostic differences between LR vs. HR and SR vs. HR MRI, respectively.

Diagnostic performance was evaluated by calculating the sensitivity, specificity, and area under the receiver operating characteristic curve (AUC). As such, grade 3 defects graded as grade 2B were recorded as false-positive findings, and grade 2B defects graded as grade 3 were recorded as false-negative findings. Kappa values of 0.41–0.60, 0.61–0.80, and 0.81–1.0 indicated moderate, substantial, and almost perfect interrater agreement, respectively. $P < 0.05$ was considered to indicate statistical significance. All computations were performed using R software (version 4.3.0).

## Results

### Participant Characteristics

The 7T MR knee paired dataset finally comprised 42 participants (mean age, 45 years ± 17 [SD]; 18 male) with 54 knees (32 left, 22 right knees). Ten participants subsequently underwent arthroscopic knee surgery. The mean interval between the MRI examination and arthroscopic knee surgery was 26 days ± 21 (minimum, 1 day; maximum, 58 days).

### Image Quality and Visibility of Anatomic Structures

The numerical results of Image Quality and Visibility of Anatomic Structures are provided in **Table 2**. The inter-reader agreement for assessing the qualitative measurements across all readers ranged from 0.620 to 0.926.

For the 7T paired data, the HR (5 [IQR, 5-5]) and SR images (5 [IQR, 5-5]) provided superior overall image quality than LR images (4 [IQR, 4-4], $P < 0.001$).

The comparisons of overall image quality between HR and SR images were insignificant ($P > 0.99$). The noise levels of SR (5 [IQR, 5-5]) and LR images (5 [IQR, 5-5]) were lower compared with those of HR images at 7T (4 [IQR, 4-4], $P < 0.001$). The HR images (5 [IQR, 5-5]) exhibited fewer Gibbs artifacts than SR (4 [IQR, 4-4]) and LR (4 [IQR, 4-4]) images ($P < 0.001$). SR images did not show fewer Gibbs artifacts compared to LR images ($P > 0.99$). For the motion artifacts, there was no significant difference across all images ($P = 0.165$). There were no image reconstruction artifacts.

The HR and SR images provided superior articular cartilage, meniscus, cruciate ligament, and collateral ligament, and tibial nerve than LR images ($P < 0.001$). However, SR and HR images showed no significant difference in the visibility of knee structure ($P \geq 0.158$) **(Figures 2-3)**. There were no indications of the introduction of hallucinations or spurious omissions of abnormalities on SR MR images.

**Frequencies of Structural Abnormalities**

**Table 3** provides the numerical results of the detection frequency, agreement, and comparison analyses. Inter-reader agreements for the detection frequencies of structural abnormalities were good or better (AC2 $\geq$ 0.738) across all datasets.

There were no statistically significant differences in detection frequencies between LR vs. HR images and SR vs. HR images for the anterior and posterior angles of medial meniscus tears (23.3% [14 of 60] for all) and lateral meniscus tears (6.7% [4 of 60] for all), and anterior and posterior cruciate ligament tears (3.3 % [2 of 60] for all, $P > 0.99$ for all). There were no medial collateral ligament tears, lateral collateral ligament tears, or patellar tendon tears. No significant differences were found in the detection frequencies of bone marrow abnormalities when comparing LR with HR images (16.2% vs. 17.6%; $P = 0.25$) or SR with HR images (17.6% for both; $P > 0.99$).

Additionally, although no significant differences were found in the detection frequencies of articular cartilage abnormalities when comparing LR with HR images ($P = 0.095$) or SR with HR images ($P = 0.406$), the intermethod agreement was higher between SR and HR images (κ = 0.95) than between LR and HR images (κ = 0.84).

**Diagnostic Performance**

We compared the diagnostic performance of LR, SR, and HR 7T knee MR images in ten participants who underwent arthroscopic surgery. Tears of the medial and lateral meniscus (20% [8 of 40]) and anterior cruciate ligament (10% [1 of 10]) were diagnosed with sensitivities of 100%, specificities of 100%, accuracies of 100%, and AUCs of 1.00 in LR, SR, and HR images. There was no incidence of posterior cruciate ligament tears in these ten participants.

The diagnostic performance for detecting cartilage defects was similar between LR, SR, and HR images ($P > .99$). Sensitivities were 62% (95% CI: 36, 82) for LR, and 69% (95% CI: 42, 87) for both SR and HR. Specificities were 98% (95% CI: 89, 99) for LR and SR, and 96% (95% CI: 86, 99) for HR. The AUC values were 0.86 (95% CI: 0.77, 0.94) for LR, 0.87 (95% CI: 0.77, 0.96) for SR, and 0.86 (95% CI: 0.77, 0.94) for HR **(Table 4, Figures 4-5)**.

**Discussion**

This study demonstrated that a deep learning superresolution algorithm (HAT) substantially enhanced the visual quality of 7T knee original low-resolution (LR) images (0.8 mm in-plane resolution) and shortened scan time. However, it did not improve radiologists' diagnostic performance for common internal derangements compared to the original images.

The superresolution (SR) images were sharper, with lower noise and more precise anatomical detail than the original LR images, confirming that the HAT model effectively performed its technical task. Radiologists assigned significantly higher image quality scores to SR images than to LR images, indicating a superior depiction of articular cartilage, menisci, ligaments, and tibial nerve in the synthetic high-resolution images. This finding is consistent with prior work by Lyu et al (21), who reported improved image quality using a deep learning superresolution approach for MR imaging.

Unlike some prior deep learning reconstruction methods(22, 23) that emphasize noise reduction and smoothing of images, our superresolution approach was designed to preserve high-frequency textural information critical for diagnosis. This strategy helped maintain subtle details that might be clinically important. For example, in one case from our study (Figure 2), a faint area of increased signal in the posterior root of the medial meniscus was barely visible on the LR image but was clearly delineated on the SR and High-resolution(HR) images as a small meniscal vessel rather than a tear. Without preserving this fine detail, the typical vascular structure could have been misinterpreted as a pathology. Similarly, the SR images improved visualization of small structures such as the tibial nerve in the popliteal region (Figure 3), revealing internal fascicular architecture that was indistinct on the LR images. These examples demonstrate that HAT-based reconstruction can enhance specific anatomical features compared to conventional imaging, while effectively avoiding the over-smoothing typically observed in other methods.

From a technical standpoint, our superresolution model prioritized maintaining authentic image texture over aggressive artifact suppression. We observed mild Gibbs artifacts in some images, which is caused by insufficient sampling of high-frequency data in the k-space domain and usually appears when the acquisition window limits data acquisition (24, 25). Notably, these artifacts were present to a similar degree on both LR and SR images and did not affect diagnostic confidence in our reader study. We chose not to apply additional filtering to remove such artifacts, reasoning that overly smoothing the images to suppress ringing might also erase or blur true pathology (23). Future refinements of the technique could integrate dedicated artifact-reduction modules, but our findings suggest that the level of artifacts in the current SR images was not significant enough to impair diagnostic accuracy. Preserving genuine tissue texture and lesion conspicuity was more critical in our approach than achieving perfectly artifact-free images.

Importantly, however, neither the synthetic SR images nor even the acquired HR (0.4-mm in-plane resolution) images at 7T led to any significant improvement in diagnostic sensitivity, specificity, or AUC values for the meniscal tears, cartilage defects, ligament injuries, and bone marrow lesions in our study when compared with the standard 0.8-mm in-plane resolution images. Furthermore, although SR images showed higher inter-method agreement with HR images (κ = 0.95) than LR images (κ = 0.84) for detecting articular cartilage defects, this difference was also not statistically significant. This lack of diagnostic performance gain suggests that the baseline 0.8-mm in-plane resolution images at 7T was already sufficient for identifying the key pathological findings in our patient cohort. The additional spatial resolution provided by SR and HR images did not alter clinical interpretations. MR detection of internal knee injuries depends on more than just pixel size. Key factors include intrinsic tissue contrast and signal changes associated with pathology(26). Once image quality surpasses a certain threshold necessary to visualize a lesion, further increases in resolution at 7T may yield diminishing diagnostic returns. Our results underscore an important principle: improvements in technical image quality do not automatically translate into better diagnostic outcomes, especially when the original imaging is already sufficient for the task at hand.

This study has several limitations. First, not all participants had a surgical reference standard; arthroscopic confirmation of findings was available in only a subset of cases, which limits the definitive assessment of diagnostic accuracy. Second, the sample size was relatively small and drawn from a single institution, which may constrain the generalizability of our results. Third, we evaluated only one MR sequence (the PD-TSE FS sequence at 7T); the benefits of superresolution might differ with other pulse sequences or field strengths, and this warrants further investigation. Fourth, we focused on the HAT model and did not compare it with different deep learning superresolution algorithms, so we cannot determine whether HAT provides superior performance relative to alternative approaches. Finally, as with any generative model, there remains a possibility of subtle image inaccuracies or hallucinated details. Although we did not encounter any obvious erroneous structures in the SR images, ongoing validation is necessary to ensure that the super-resolved images accurately represent the true anatomy in diverse clinical cases.

## Conclusion

Deep learning superresolution substantially improved the spatial resolution and perceived image quality of 7T knee MR images. Although diagnostic performance was comparable to standard low-resolution imaging, the technique shows promise for enhancing anatomic visualization and may provide added value in applications requiring fine structural detail.

TABLE

Table 1. PD-TSE FS 7T knee MR imaging sequence parameters for high-resolution and low-resolution acquisitions

| Sequence | Higher resolution | | | Lower resolution | | |
|---|---|---|---|---|---|---|
| | SAG | COG | TRA | SAG | COG | TRA |
| Repetition time (ms) | 4500 | 4500 | 4500 | 4500 | 4500 | 4500 |
| Echo time (ms) | 31 | 31 | 31 | 36 | 36 | 36 |
| Flip angle (°) | 150 | 140 | 150 | 150 | 140 | 150 |
| Average | 1 | 1 | 1 | 1 | 1 | 1 |
| Accel. factor | 2 | 2 | 2 | 2 | 2 | 2 |
| Bandwidth | 685 | 685 | 685 | 685 | 685 | 685 |
| Field of View (mm) | 160×160 | 160×160 | 160×160 | 160×160 | 160×160 | 160×160 |
| Matrix | 384×384 | 384×384 | 384×384 | 192×192 | 192×192 | 192×192 |
| Resolution (mm$^3$) | 0.4×0.4×2 | 0.4×0.4×2 | 0.4×0.4×2 | 0.8×0.8×2 | 0.8×0.8×2 | 0.8×0.8×2 |
| Slice | 36 | 35 | 40 | 36 | 35 | 40 |
| Time | 4min | 4min | 4min | 2min26s | 2min26s | 2min35s |

Table 2: Image quality scores for low-resolution, super-resolution, and high-resolution 7T knee MR images.

| Sequence | LR Median (IQR) | SR Median (IQR) | HR Median (IQR) | Interreader Agreement* | P Values† LR vs SR | LR vs HR | SR vs HR |
|---|---|---|---|---|---|---|---|
| Image quality | 4 [4,4] | 5 [5,5] | 5 [5,5] | 0.845 [0.719, 0.935] | < 0.001 | < 0.001 | > 0.99 |
| Noise | 5 [5,5] | 5 [5,5] | 4 [4,4] | 0.926 [0.831, 1.000] | > 0.99 | < 0.001 | < 0.001 |
| Motion Artifacts | 5 [4.8,5] | 5 [5,5] | 5 [5,5] | 0.684 [0.435, 0.870] | Ns | Ns | Ns |
| Gibbs Artifacts | 4 [4,4] | 4 [4,4] | 5 [5,5] | 0.668 [0.482, 0.816] | > 0.99 | < 0.001 | < 0.001 |
| Articular cartilage | 4 [4,4] | 5 [5,5] | 5 [5,5] | 0.809 [0.678, 0.927] | < 0.001 | < 0.001 | > 0.99 |
| Meniscus | 4 [4,4] | 5 [5,5] | 5 [5,5] | 0.832 [0.703, 0.932] | < 0.001 | < 0.001 | > 0.99 |
| Cruciate ligament | 4 [4,4] | 5 [5,5] | 5 [5,5] | 0.924 [0.820, 1.000] | < 0.001 | < 0.001 | > 0.99 |
| Collateral liagment | 4 [4,4] | 5 [5,5] | 5 [5,5] | 0.783 [0.652, 0.900] | < 0.001 | < 0.001 | > 0.99 |
| Tibial nerve | 3 [3,3] | 3.5 [3.5,4] | 4 [4,4] | 0.620 [0.453, 0.768] | < 0.001 | < 0.001 | 0.158 |

Data is the median (IQR) of the average values derived from the five grades assigned by three readers. 'Ns' indicates not significant, with the P values of 0.165 for Motion Artifacts. LR: Lower resolution. SR: Super-resolution. HR: High-resolution.

* Data are Gwet AC2 values with 95% CIs in parentheses.  † The Friedman test was

followed by a pairwise post hoc Wilcoxon test with the Bonferroni-Holm correction.

$P < 0.05$ means the difference is significant.

Table 3: Detection frequencies of knee structural abnormalities on low-resolution, super-resolution, and high-resolution 7T MR images.

| Structural Abnormalities | Reader 1 | Reader 2 | Reader 3 | Interreader Agreement* | Consensus Score | LR vs HR | | SR vs HR | |
|---|---|---|---|---|---|---|---|---|---|
| | | | | | | Kappa† | P value§ | Kappa† | P value§ |
| The anterior and posterior angles of Medial meniscus tears | | | | | | 1 | > 0.99 | 1 | > 0.99 |
| LR | 14/60 | 14/60 | 14/60 | 1 | 14/60 (23.3%) | | | | |
| SR | 14/60 | 14/60 | 14/60 | 1 | 14/60 (23.3%) | | | | |
| HR | 14/60 | 14/60 | 14/60 | 1 | 14/60 (23.3%) | | | | |
| The anterior and posterior angles of Lateral meniscus tears | | | | | | 1 | > 0.99 | 1 | > 0.99 |
| LR | 4/60 | 5/60 | 5/60 | 0.845[0.840, 0.850] | 4/60 (6.7%) | | | | |
| SR | 4/60 | 4/60 | 4/60 | 1 | 4/60 (6.7%) | | | | |
| HR | 4/60 | 4/60 | 4/60 | 1 | 4/60 (6.7%) | | | | |
| Anterior and Posterior cruciate ligament tears | | | | | | 1 | > 0.99 | 1 | > 0.99 |
| LR | 2/60 | 2/60 | 4/60 | 0.738[0.734, 0.743] | 2/60 (3.3%) | | | | |
| SR | 2/60 | 2/60 | 3/60 | 0.738[0.732, 0.745] | 2/60 (3.3%) | | | | |
| HR | 2/60 | 3/60 | 2/60 | 0.851[0.847, 0.856] | 2/60 (3.3%) | | | | |
| Bone marrow abnormalities | | | | | | 0.95[0.88,0.99] | 0.25 | 1 | > 0.99 |
| LR | 34/210 | 36/210 | 35/210 | 0.943[0.940, 0.945] | 34/210 (16.2%) | | | | |
| SR | 37/210 | 37/210 | 38/210 | 0.967[0.965, 0.970] | 37/210 (17.6%) | | | | |
| HR | 37/210 | 38/210 | 38/210 | 0.957[0.954, 0.959] | 37/210 (17.6%) | | | | |
| Articular cartilage defects | | | | | | 0.84[0.74, 0.91] | 0.095 | 0.95[0.90, 0.99] | 0.406 |
| LR-Grade 0 | 135/180 | 131/180 | 135/180 | 0.905[0.903, 0.906] | 135/180 (75%) | | | | |
| LR-Grade 1 | 7/180 | 10/180 | 4/180 | | 7/180 (3.9%) | | | | |
| LR-Grade 2A | 16/180 | 16/180 | 18/180 | | 16/180 (8.9%) | | | | |
| LR-Grade 2B | 12/180 | 12/180 | 12/180 | | 12/180 (6.7%) | | | | |
| LR-Grade 3 | 10/180 | 10/180 | 10/180 | | 10/180 (5.6%) | | | | |

| | | | | | | | | | | |
|---|---|---|---|---|---|---|---|---|---|---|
| SR-Grade 0 | 132/80 | 128/180 | 134/180 | 0.917[0.916, 0.919] | 132/180(73.3%) | | | | | |
| SR-Grade 1 | 11/180 | 11/180 | 8/180 | | 11/180 (6.1%) | | | | | |
| SR-Grade 2A | 16/180 | 22/180 | 17/180 | | 17/180 (9.4%) | | | | | |
| SR-Grade 2B | 10/180 | 9/180 | 11/180 | | 10/180 (5.6%) | | | | | |
| SR-Grade 3 | 11/180 | 10/180 | 10/180 | | 10/180 (5.6%) | | | | | |
| HR-Grade 0 | 130/180 | 130/180 | 133/180 | 0.951[0.949, 0.952] | 130/180(72.2%) | | | | | |
| HR-Grade 1 | 13/180 | 13/180 | 9/180 | | 13/180 (7.2%) | | | | | |
| HR-Grade 2A | 18/180 | 19/180 | 17/180 | | 18/180 (10%) | | | | | |
| HR-Grade 2B | 9/180 | 8/180 | 11/180 | | 9/180 (5%) | | | | | |
| HR-Grade 3 | 10/180 | 10/180 | 10/180 | | 10/180 (5.6%) | | | | | |

Note.—Unless otherwise indicated, data are numerators/denominators. The frequencies of bone marrow abnormalities are based on seven bone compartments and the frequencies of articular cartilage defects are based on six cartilage compartments per subject. LR: Lower resolution. SR: Super-resolution. HR: High-resolution.

* Data are Gwet AC2 values with 95% CIs in parentheses.

† κ values with 95% Cis were computed using Cohen's kappa test to quantify intermethod agreement for diagnosing structural abnormalities between LR and HR or SR and HR MRI.

§ P values were computed using McNemar's test to compare diagnosis differences between LR and HR or SR and HR MRI.

Table 4: Diagnostic performance for detecting articular cartilage lesions on low-resolution, superresolution, and high-resolution 7T knee MR images.

| Images | Prevalence at Arthroscopy (%) | Frequency at MRI (%) | No. of Findings | | | | Sensitivity (%) | Specificity (%) | AUC |
|---|---|---|---|---|---|---|---|---|---|
| | | | TN | TP | FN | FP | | | |
| LR | | | 46 | 8 | 1 | 5 | 62 [36,82] | 98 [89,99] | 0.86[0.77,0.94] |
| Grade 0 | 78 (47/60) | 82 (49/60) | 46 | | | | | | |
| Grade 1 | 3 (2/60) | 2 (1/60) | | 1 | | | | | |
| Grade 2A | 8 (5/60) | 8 (5/60) | | 3 | | | | | |
| Grade 2B | 2 (1/60) | 2 (1/60) | | 0 | | | | | |
| Grade 3 | 8 (5/60) | 7 (4/60) | | 4 | | | | | |
| SR | | | 46 | 9 | 1 | 4 | 69 [42,87] | 98 [89,99] | 0.87[0.77,0.96] |
| Grade 0 | 78 (47/60) | 82 (49/60) | 46 | | | | | | (0.77,0.96) |
| Grade 1 | 3 (2/60) | 2 (1/60) | | 1 | | | | | |
| Grade 2A | 8 (5/60) | 7 (4/60) | | 3 | | | | | |
| Grade 2B | 2 (1/60) | 2 (1/60) | | 0 | | | | | |
| Grade 3 | 8 (5/60) | 8 (5/60) | | 5 | | | | | |
| HR | | | 45 | 9 | 2 | 4 | 69 [42, 87] | 96 [86, 99] | 0.86[0.77,0.94] |
| Grade 0 | 78 (47/60) | 80 (48/60) | 45 | | | | | | (0.77,0.96) |
| Grade 1 | 3 (2/60) | 2 (1/60) | | 1 | | | | | |
| Grade 2A | 8 (5/60) | 8 (5/60) | | 3 | | | | | |
| Grade 2B | 2 (1/60) | 2 (1/60) | | 0 | | | | | |
| Grade 3 | 8 (5/60) | 8 (5/60) | | 5 | | | | | |

Note: Prevalence and frequency values are given as percentages, with numerators and denominators in parentheses. The sensitivity, specificity, and AUC are given as

percentages, with 95% CIs in parentheses. The five grades of articular cartilage(grade 0, grade 1, grade 2A, grade 2B, and grade 3) were retained for the diagnostic performance analysis. As such, grade 3 defects graded as grade 2B were recorded as false-positive findings, and grade 2B defects graded as grade 3 were recorded as false-negative findings.

LR: Lower resolution. SR: Super-resolution. HR: High-resolution. AUC: area under the curve. FN: false-negative finding. FP: false-positive finding. TN, true-negative finding. TP, true-positive finding.

FIGURES

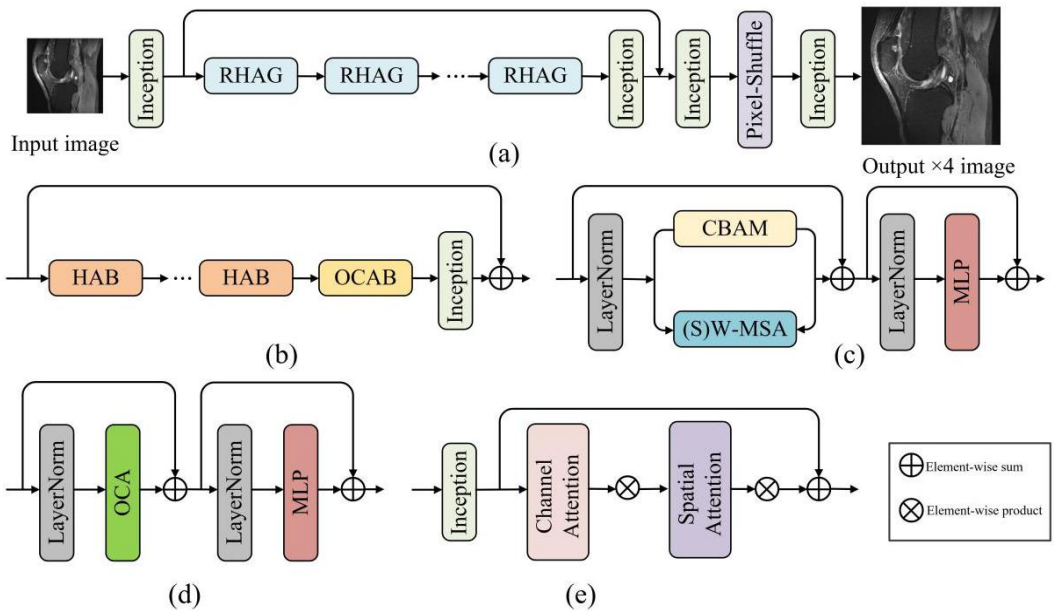

**Figure 1.** The framework of the proposed method. Diagram of the Hybrid Attention Transformer (HAT) network architecture for superresolution. (a) Overall HAT model architecture. (b) Structure of a Residual Hybrid Attention Group (RHAG). (c) Structure of a Hybrid Attention Block (HAB). (d) Structure of an Overlapping Cross-Attention Block (OCAB). (e) Structure of a Convolutional Block Attention Module (CBAM).

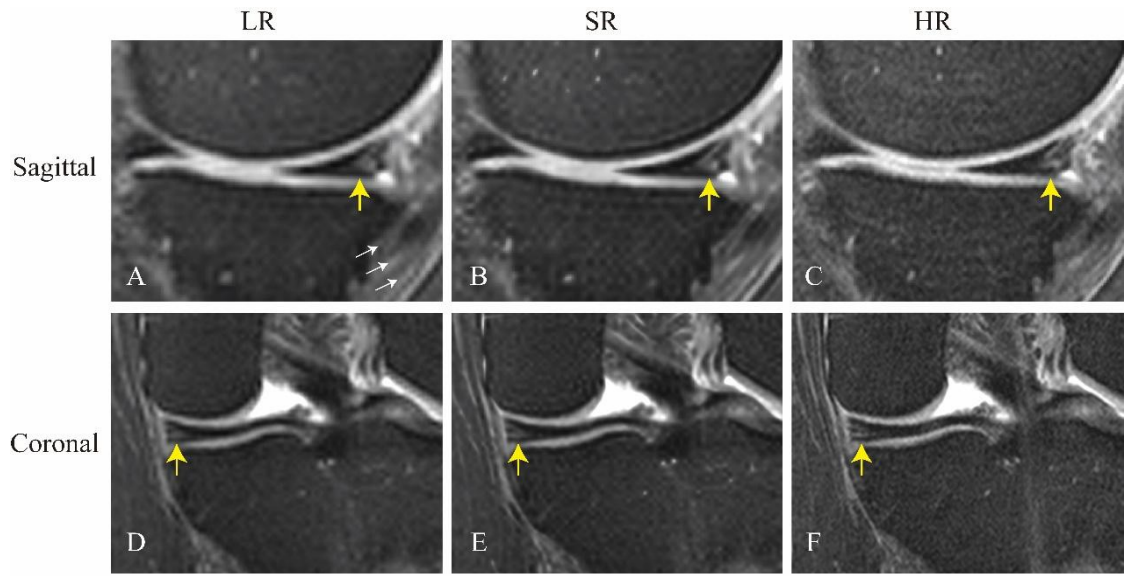

**Figure 2:** Sagittal (A–C) and coronal (D–F) PD-TSE FS 7T MR images of the left knee in a 42-year-old woman with knee pain. Low-resolution (LR) images (A, D; 0.8 mm in-plane) show an indistinct hyperintense area in the posterior root of the medial meniscus (yellow arrow). The superresolution (SR) images (B, E) and high-resolution (HR) images (C, F; 0.4 mm in-plane) delineate this finding as a small meniscal vessel rather than a tear. Gibbs artifacts, characterized by parallel lines of high and low signal intensity, are indicated by white arrows in image A. (LR = low resolution, SR = superresolution, HR = high resolution).

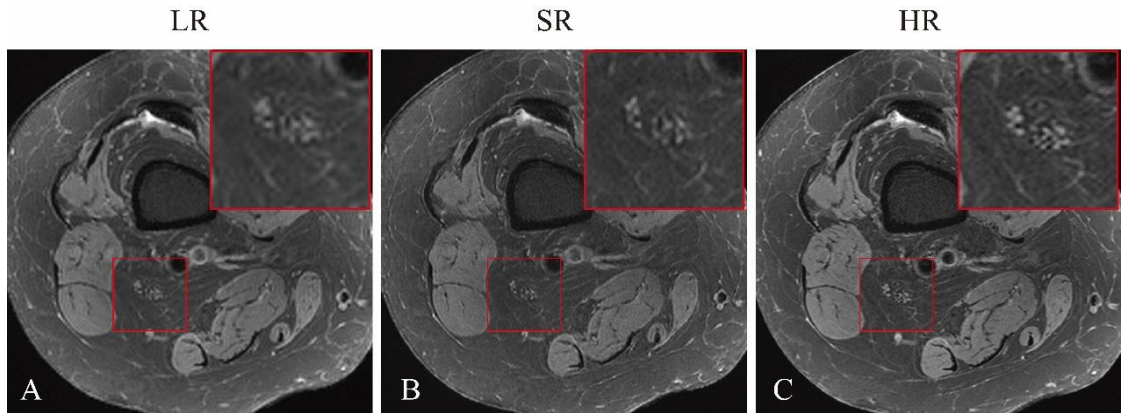

**Figure 3:** Axial PD-TSE FS 7T MR images of the right knee in a 45-year-old woman with knee pain. (A) Low-resolution image (0.8 mm in-plane), (B) superresolution image, and (C) high-resolution image (0.4 mm in-plane). The SR image (B) provides improved visualization of the the sciatic nerve bifurcation (into the tibial and common fibular nerves) and its internal fascicular detail compared with the LR image (A). Red boxes mean the close-up of nerves. (LR = low resolution, SR = superresolution, HR = high resolution).

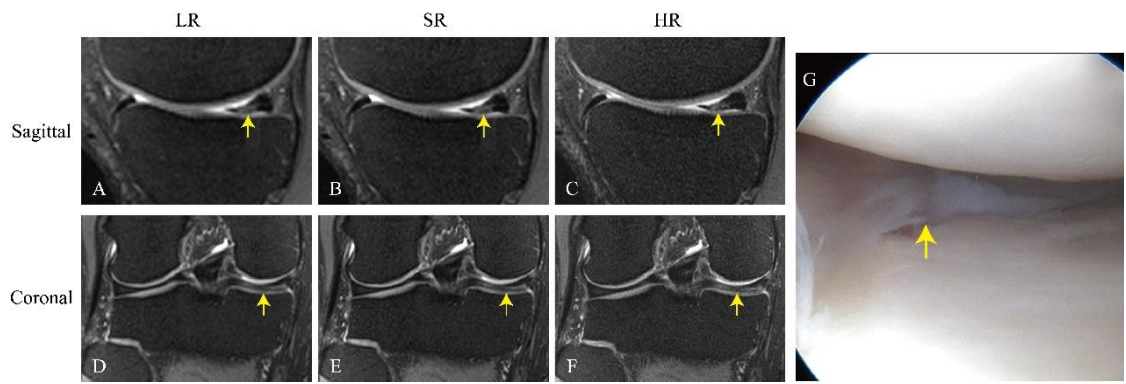

**Figure 4:** Sagittal (A–C) and coronal (D–F) PD-TSE FS 7T MR images of the right knee in a 42-year-old man with knee pain. LR images (A, D), SR images (B, E), and HR images (C, F) depict a posterior root tear of the medial meniscus (yellow arrow). An arthroscopic photograph (G) confirms the meniscal tear (arrows). (LR = low resolution, SR = superresolution, HR = high resolution).

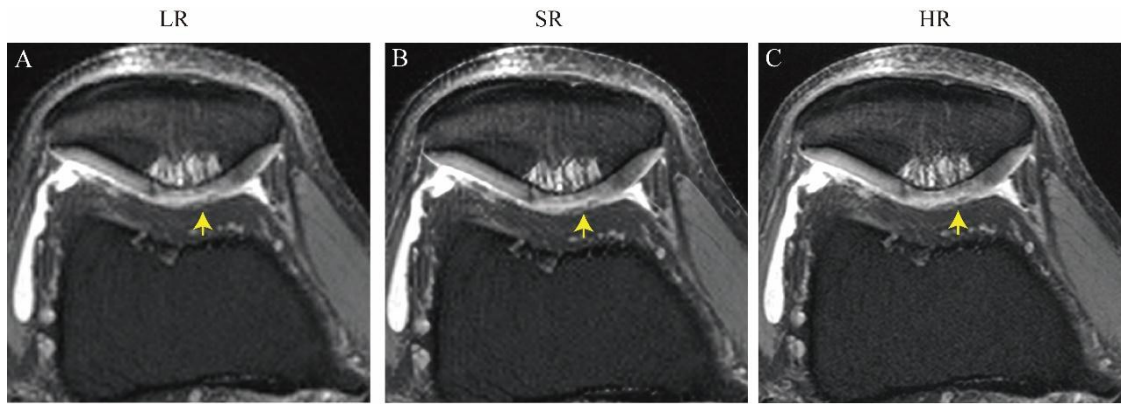

**Figure 5:** Axial PD-TSE FS 7T MR images (magnified) of the right knee in a 53-year-old man with knee pain. (A) Low-resolution image, (B) uperresolution image, and (C) high-resolution image. All three images depict an incidental cartilage lesion (yellow arrows), which remains clearly visible even on the LR image. (LR = low resolution, SR = superresolution, HR = high resolution).